\documentclass[a4paper,11pt]{article}
\usepackage{pos}

\title{Using conditional variational autoencoders to generate images from atmospheric Cherenkov telescopes}
\ShortTitle{Using CVAEs to generate images from IACTs}

\author*[a]{Stanislav Polyakov}
\author[a]{Alexander Kryukov}
\author[a]{Andrey Demichev}
\author[a]{Julia Dubenskaya}
\author[b]{Elizaveta Gres}
\author[c]{Anna Vlaskina}

\affiliation[a]{Skobeltsyn Institute of Nuclear Physics, Lomonosov Moscow State University,\protect\\
 1(2) Leninskie gory, GSP-1, Moscow 119991, Russian Federation}
\affiliation[b]{Applied Physics Institute of Irkutsk State University, \protect\\
20, Gagarin boulevard, Irkutsk, 664003, Russian Federation}
\affiliation[c]{Lomonosov Moscow State University, \protect\\
 1(2) Leninskie gory, GSP-1, Moscow 119991, Russian Federation}

\emailAdd{s.p.polyakov@gmail.com}
\emailAdd{kryukov@theory.sinp.msu.ru}
\emailAdd{demichev@theory.sinp.msu.ru}
\emailAdd{jdubenskaya@gmail.com}
\emailAdd{greseo@mail.ru}
\emailAdd{nina.vankalas@gmail.com}

\abstract{High-energy particles hitting the upper atmosphere of the Earth produce extensive air showers that can be detected from the ground level using imaging atmospheric Cherenkov telescopes. The images recorded by Cherenkov telescopes can be analyzed to separate gamma-ray events from the background hadron events. Many of the methods of analysis require simulation of massive amounts of events and the corresponding images by the Monte Carlo method. However, Monte Carlo simulation is computationally expensive. The data simulated by the Monte Carlo method can be augmented by images generated using faster machine learning methods such as generative adversarial networks or conditional variational autoencoders. We use a conditional variational autoencoder to generate images of gamma events from a Cherenkov telescope of the TAIGA experiment. The variational autoencoder is trained on a set of Monte Carlo events with the image size, or the sum of the amplitudes of the pixels, used as the conditional parameter. We used the trained variational autoencoder to generate new images with the same distribution of the conditional parameter as the size distribution of the Monte Carlo-simulated images of gamma events. The generated images are similar to the Monte Carlo images: a classifier neural network trained on gamma and proton events assigns them the average gamma score 0.984, with less than 3\% of the events being assigned the gamma score below 0.999. At the same time, the sizes of the generated images do not match the conditional parameter used in their generation, with the average error 0.33.}

\FullConference{%
  *** The 6th International Workshop on Deep Learning in Computational Physics (DLCP2022) ***\\
  *** 6-8 July 2022 ***\\
  *** JINR, Dubna, Russia ***
}

\begin{document}
\maketitle

\section{Introduction}

Extensive air showers produced by high-energy cosmic rays and gamma quanta hitting the upper atmosphere of the Earth can be detected using imaging atmospheric Cherenkov telescopes (IACTs). The images from Cherenkov telescopes are used for analysis of the events, most importantly, for identifying the type of the primary particle. Gamma quanta are of interest in gamma-ray astronomy, but background cosmic ray events are at least three orders of magnitude more common so the classification tools need to have very high specificity. A widely used approach to the analysis of the Cherenkov telescope images is comparing them with simulated images of air showers with known parameters of the primary particles generated by Monte Carlo method. Thanks to recent advancements in machine learning, this approach is particularly attractive because the generated images can be used as the training data for supervised learning. Machine learning methods often show better results with large numbers of training examples, but Monte Carlo simulation has high computational costs. It is possible to augment the training data using a smaller data set to generate more images without the Monte Carlo method, for example, by using generative adversarial networks (GANs) \cite{Goodfellow14}. This approach is not limited to gamma-ray astronomy: for example, it was used by the ATLAS experiment at the Large Hadron Collider \cite{ATLAS20}. 

We use machine learning tools to generate additional images for Cherenkov telescopes of the TAIGA experiment \cite{Budnev22}. In \cite{Dubenskaya22a, Dubenskaya22b} generative adversarial networks were trained to generate TAIGA IACT images of gamma and proton events. In this work, we present preliminary results of using conditional variational autoencoders to generate images of gamma events. 

A variational autoencoder (VAE) \cite{Kingma13} is an artificial neural network that consists of two components: an encoder and a decoder. They are trained together to learn a distribution of a set of latent variables corresponding to the distribution of the training data, that can be used to reconstruct the input data with minimal losses. VAEs can be used to generate new data points from the same distribution as the training set. They have a potential advantage over GANs in allowing for easier interpretation of the variables used to generate new data points \cite{Mathieu19}. Conditional variational autoencoders (CVAEs) \cite{Sohn15} add explicitly known parameters to the latent parameters. The conditional parameters can be either discrete like the type of the primary particle, or continuous like the energy of the event. It is also possible to enforce a condition as a constraint if it can be explicitly calculated from the data (see \cite{Zeng19} for constrained GANs).

\section{Methods}

\begin{figure}[h!]
\centering
\includegraphics[scale=0.5]{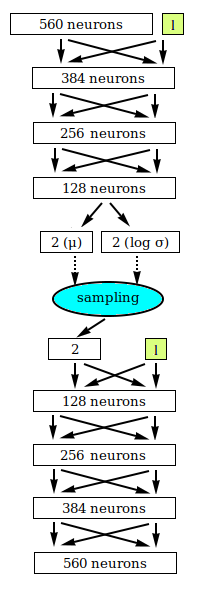}
\smallskip
\caption{The architecture of the conditional variational autoencoder for 560 pixel images. The conditional parameter added to the inputs is highlighted in yellow.}
\label{fig:architecture}
\end{figure}

The architecture of the conditional variational autoencoder we use is presented in Figure~\ref{fig:architecture}. The amplitudes $A$ of the 560 pixels of the input image are scaled using the function
$$
\frac{\log(1+A)}{\log(1+A_{max})}
$$
where $A_{max}$ is the amplitude of the brightest pixel in the training set. The encoder and the decoder of the CVAE both have three fully connected hidden layers with 128, 256, and 384 neurons (reverse order for the encoder). The encoder output is two 2-dimensional vectors which are interpreted as the expected values and the logarithms of variance of the hidden parameters. A~conditional variable is added to the input of both the encoder and the decoder. The logarithm of the image size (the sum of the amplitudes of the pixels) is used as the condition.

We used a set of 39443 images of gamma events with the energy between 1.5 TeV and 60 TeV generated by a Monte Carlo simulation program CORSIKA \cite{CORSIKA} for the Cherenkov telescopes of the TAIGA experiment. A~cleaning procedure \cite{Postnikov19b} was applied to the images, excluding all image pixels that were not a part of a pixel cluster with at least one pixel with the amplitude above the core threshold value of 6 photoelectrons, and at least one neighbor pixel with the amplitude above the neighbor threshold value of 3 photoelectrons. The resulting images had sizes ranging from 23 to 6064 photoelectrons. Out of these images, a training set of 29568 was randomly selected, and the remaining images were used as the test set.

The variational autoencoder was implemented in TensorFlow \cite{TensorFlow} and trained for 5000 epochs on the training set using the TensorFlow implementation of the Adam optimizer \cite{Kingma14}.

\section{Results}

\begin{figure}[h!]
\centering
\includegraphics[scale=0.08]{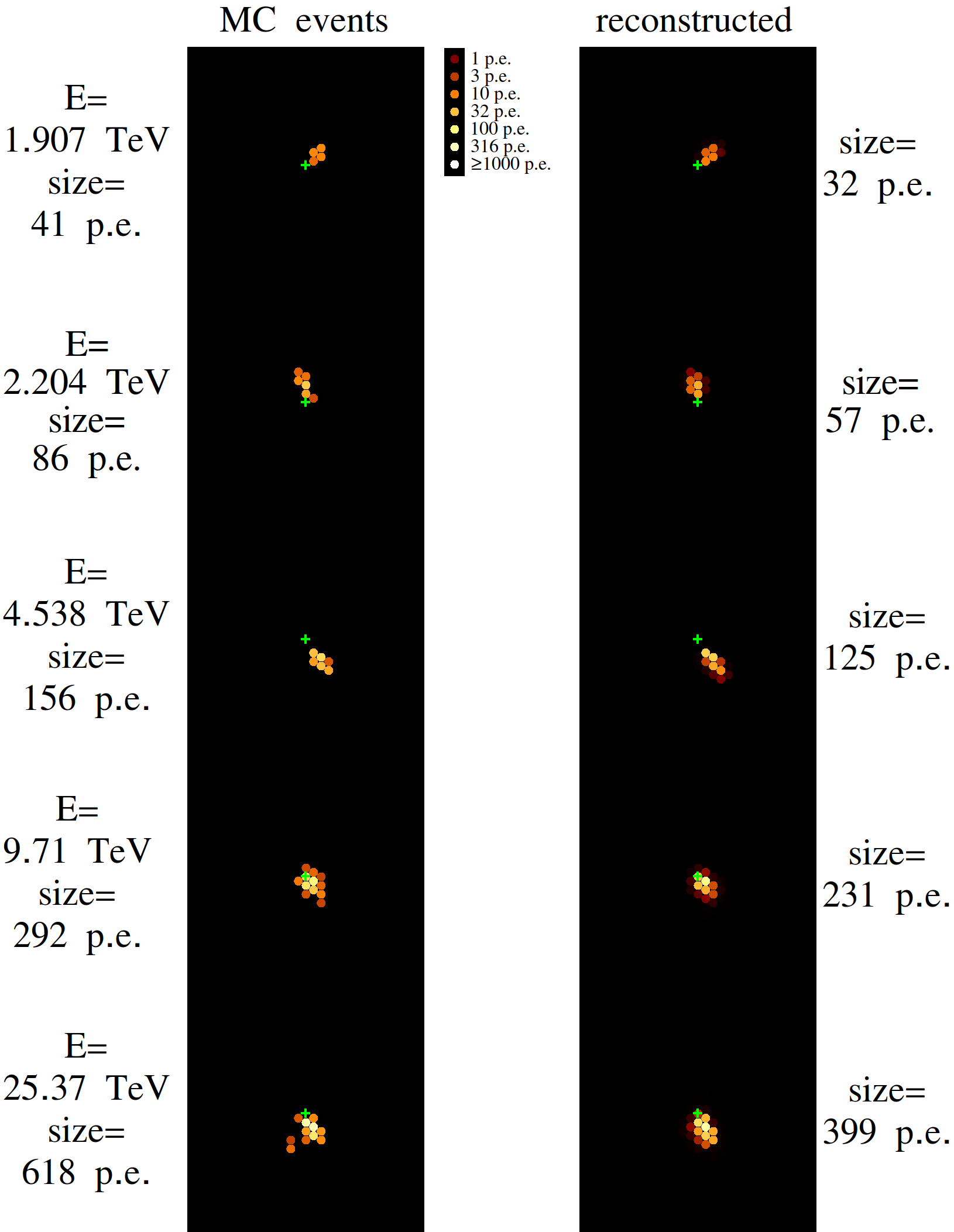}
\smallskip
\caption{The comparison of five original Monte Carlo events (left) with their reconstructions by the conditional variational autoencoder (CVAE) (right). The centers of the cameras are marked by cross hairs. P.e.~is short for photoelectrons. (The color version of the figure can be found in the online version of the paper on the journal’s website.)}
\label{fig:mc_vs_cvae}
\end{figure}

\begin{figure}[h!]
\centering
\includegraphics[scale=0.08]{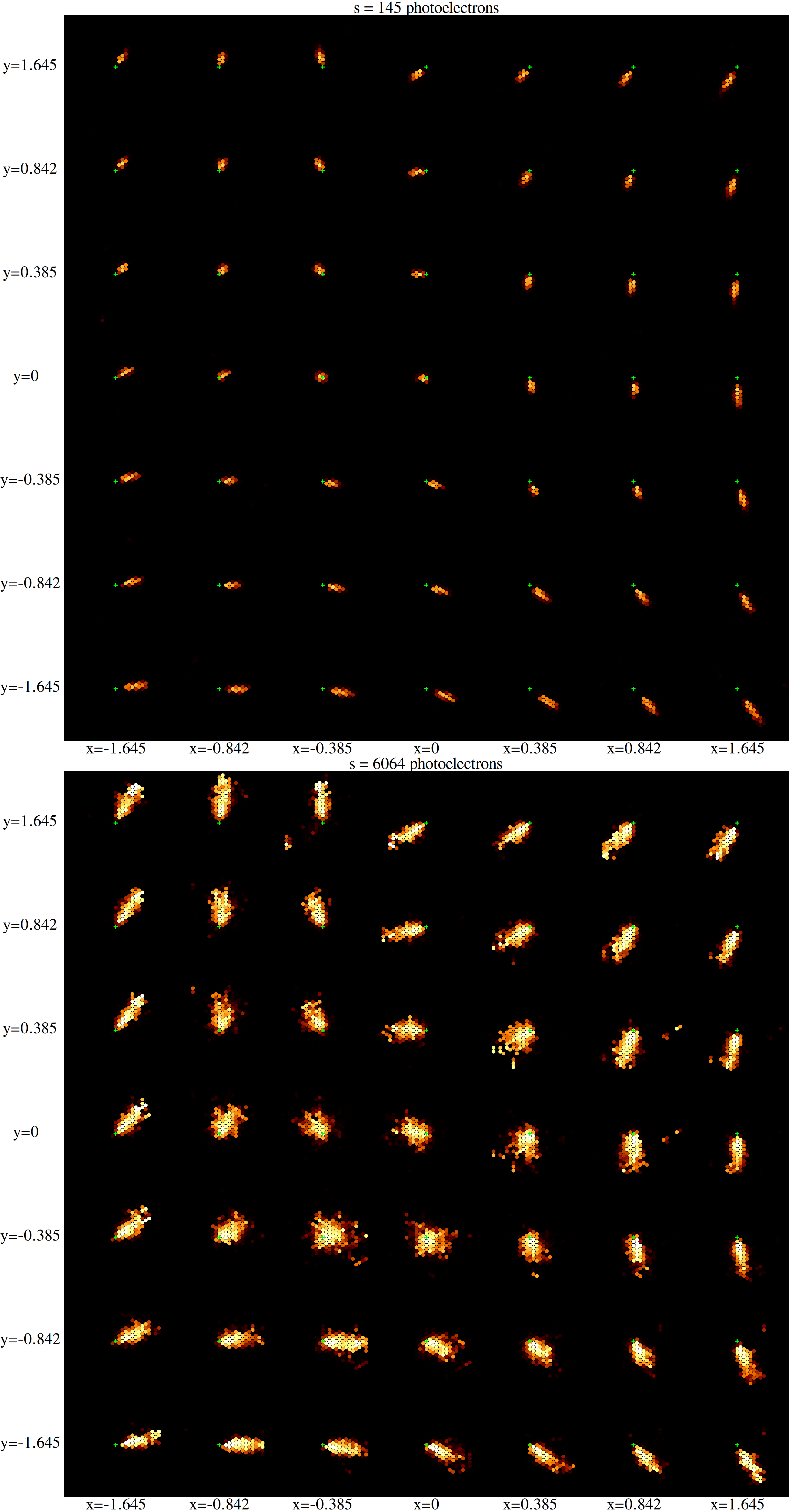}
\smallskip
\caption{Images generated by the CVAE decoder from the code (x,y) using the conditional parameter $s$.}
\label{fig:bd3_spread_s_145_and_6064}
\end{figure}

\begin{figure}[h!]
\centering
\includegraphics[scale=0.08]{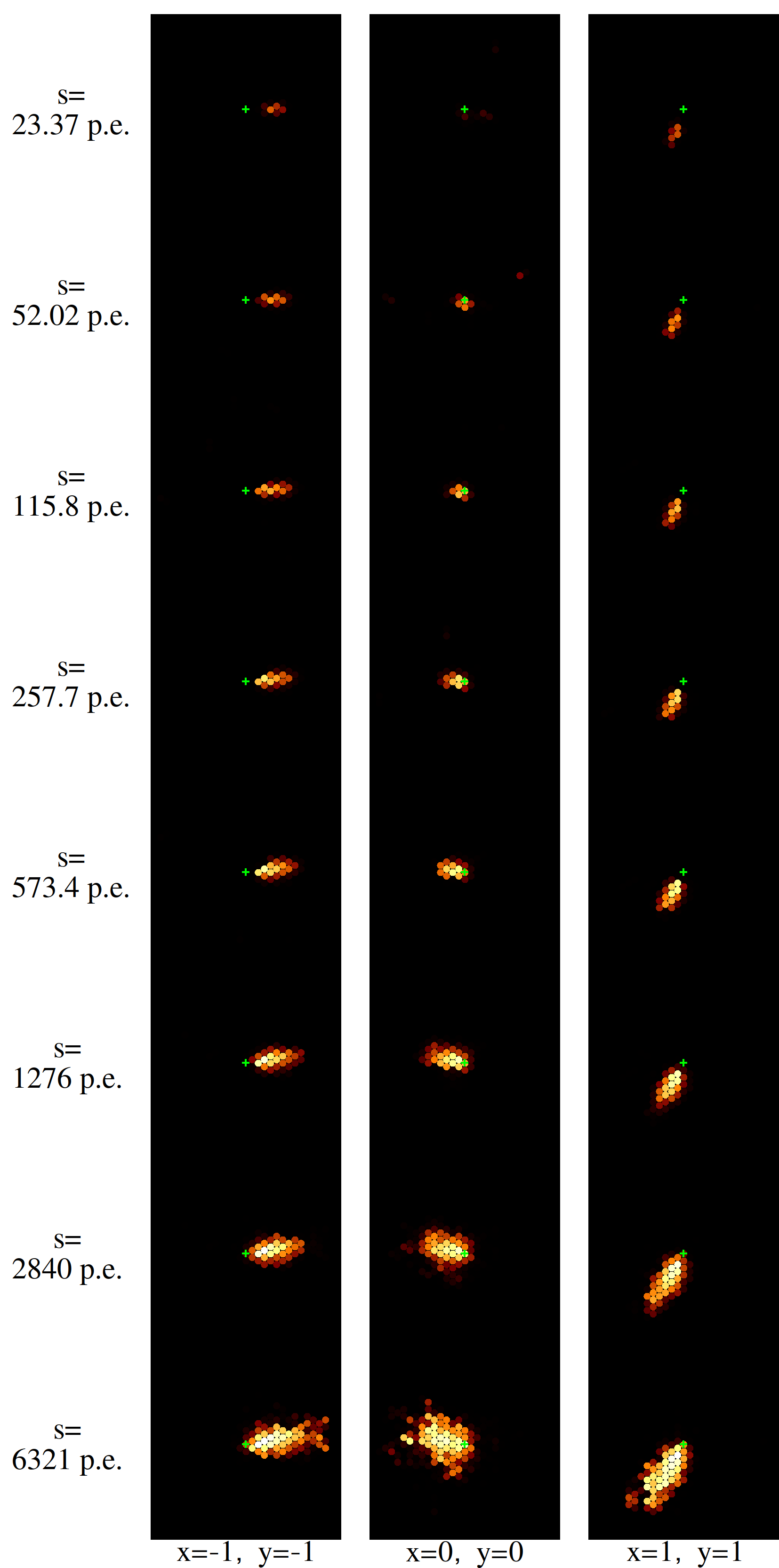}
\smallskip
\caption{Images generated by the CVAE decoder from given inputs. P.e.~is short for photoelectrons.}
\label{fig:bd3_size_series}
\end{figure}

Figure~\ref{fig:mc_vs_cvae} shows five of the original Monte Carlo events on the left side, and the respective images reconstructed by the variational autoencoder on the right side. The examples of images generated by a the trained variational autoencoder are shown in Figure~\ref{fig:bd3_spread_s_145_and_6064}. The values of the code variables $x$ and $y$ correspond to the cumulative probabilities $0.05, 0.2, 0.35, 0.5, 0.65, 0.8, 0.95$ of the normal distribution $\mathcal{N}(0,1)$, the values of the conditional parameter $s$ correspond to the median and the maximum sizes of the Monte Carlo images: 145 photoelectrons and 6064 photoelectrons, respectively.
Figure~\ref{fig:bd3_size_series} shows the changes to the images generated from the same code when the conditional parameter $s$ is varied. 

\begin{figure}[h!]
\centering
\includegraphics[scale=0.5]{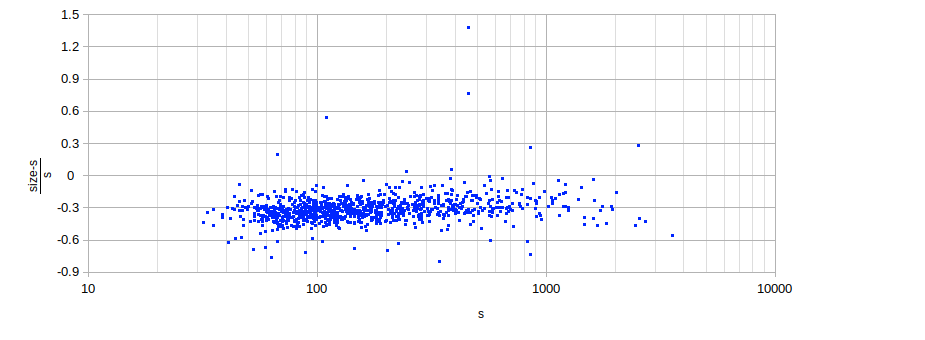}
\smallskip
\caption{The relative size shift by the CVAE: $s$ is the value of the conditional parameter and size is the sum of the pixel amplitudes in the generated image.}
\label{fig:size_shift}
\end{figure}

The sizes of the images generated by the conditional variational autoencoder tend to be smaller than the values of the conditional parameter $s$. 
For a sample of events generated with the code $(x, y)$ sampled from the normal distribution $\mathcal{N}(0,1)$ and the conditional parameter $s$ having the same distribution as the size of the Monte Carlo events, the average relative error $|\frac{size-s}{s}|$ is 0.33. Figure~\ref{fig:size_shift} shows the relative size shift for a similar sample of 1000 events. This discrepancy can be corrected by turning the conditional parameter into a constraint. However, a more simple approach is to increase the parameter $s$ by a fixed factor to get images close to the desired size, because the size shift is relatively consistent: the correlation coefficient of the parameter $s$ and the size of the generated images in this sample is 0.946.

\begin{figure}[h!]
\centering
\includegraphics[scale=0.5]{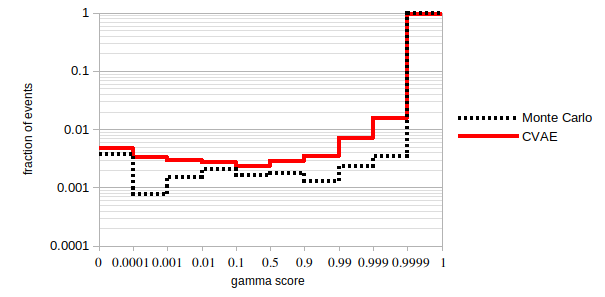}
\smallskip
\caption{The distribution of the gamma scores by the classifier neural network for Monte Carlo events and the CVAE-generated events.}
\label{fig:gamma_score}
\end{figure}

Other than the diminished size, the images generated by the CVAE generally superficially look similar to the images of Monte Carlo-simulated gamma events: they have approximately elliptical shape pointing towards the center of the camera. This is confirmed by a classifier neural network \cite{Postnikov19a} trained on the same training set plus 22751 proton events: 97.1\% of the images are assigned the gamma score above 0.999, and the average gamma for the sample is 0.984, compared to 0.99 assigned to the Monte Carlo gamma events not used in the training set of the classifier. The distribution of the gamma scores is shown in Figure~\ref{fig:gamma_score}.

\section{Conclusion and further research}

We have trained conditional variational autoencoders on a set of TAIGA Cherenkov telescope images of Monte Carlo-simulated gamma events. The resulting autoencoder can be used to generate new images that are similar to the gamma event images: in particular, a classifier neural network assigns them the average gamma score 0.984. The conditional parameter $s$ was based on the sizes of the training images, however, the size of the generated images does not match the values used as the decoder input. The average relative error is 0.33, with most of the events having lower size than the parameter $s$. 

The possible directions of further research include enforcing a constraint on a variational autoencoder to make the generated image parameters match those used as conditions and using convolutional layers. We also plan to use CVAE-generated images to augment training sets for the neural networks used for image analysis, in particular, event type classifiers.

\acknowledgments{The work was supported by the Russian Science Foundation (grant No. 22-21-00442).}

\bibliographystyle{JHEP}
\bibliography{bib_polyakov_s}

\end{document}